\documentclass[aps,prd,reprint,superscriptaddress,nofootinbib, longbibliography]{revtex4-2}

\usepackage{amssymb}
\usepackage{amsmath}
\usepackage[utf8]{inputenc}
\usepackage{graphicx}
\usepackage{cancel}
\usepackage{bbm}
\usepackage{color}
\usepackage{lineno}
\usepackage{dcolumn}   % needed for some tables
\usepackage{bm}        % for math
\usepackage{amssymb}   % for math
\usepackage{amsmath,amssymb}
\usepackage{comment}
\usepackage[dvipsnames]{xcolor}
\usepackage[colorlinks = true,
            linkcolor  = RoyalBlue,
            urlcolor   = RoyalBlue,
            citecolor  = RoyalBlue]{hyperref}
            
\usepackage[utf8]{inputenc}
\usepackage{amsmath}
\usepackage{graphicx}
\usepackage{color}
\newcommand{\nineH}        {$\sqrt{s}~=~0.9$~Te\kern-.1emV\xspace}
\newcommand{\seven}        {$\sqrt{s}~=~7$~Te\kern-.1emV\xspace}
\newcommand{\twoH}         {$\sqrt{s}~=~0.2$~Te\kern-.1emV\xspace}
\newcommand{\twosevensix}  {$\sqrt{s}~=~2.76$~Te\kern-.1emV\xspace}
\newcommand{\five}         {$\sqrt{s}~=~5.02$~Te\kern-.1emV\xspace}
\newcommand{\twosevensixnn}{$\sqrt{s_{\mathrm{NN}}}~=~2.76$~Te\kern-.1emV\xspace}
\newcommand{\fivenn}       {$\sqrt{s_{\mathrm{NN}}}~=~5.02$~Te\kern-.1emV\xspace}

\begin{document}
\newcommand{\trento}{$\mathrm{T_RENTo}$~}
\title{Fluid dynamics of charm quarks in the quark--gluon plasma}

\author{F.~Capellino}
\email[]{f.capellino@gsi.de}
\affiliation{Physikalisches Institut, Universit{\"a}t Heidelberg, 69120 Heidelberg, Germany}
\affiliation{GSI Helmholtzzentrum f{\"u}r Schwerionenforschung, 64291 Darmstadt, Germany}

\author{A.~Dubla}
%\email[]{a.dubla@cern.ch}
\affiliation{GSI Helmholtzzentrum f{\"u}r Schwerionenforschung, 64291 Darmstadt, Germany}

\author{S.~Floerchinger}
%\email[]{stefan.floerchinger@uni-jena.de}
\affiliation{Theoretisch-Physikalisches Institut
Friedrich-Schiller-Universität Jena, 07743 Jena, Germany} 

\author{E.~Grossi}
%\email[]{eduardo.grossi@unifi.it}
\affiliation{Dipartimento di Fisica, Universit\`a di Firenze and INFN Sezione di Firenze,
50019 Sesto Fiorentino, Italy}

\author{A.~Kirchner}
%\email[]{kirchner@thphys.uni-heidelberg.de }
\affiliation{Institut f\"ur Theoretische Physik Heidelberg, 69120 Heidelberg, Germany} 

\author{S.~Masciocchi}
%\email[]{s.masciocchi@gsi.de}
\affiliation{Physikalisches Institut, Universit{\"a}t Heidelberg, 69120 Heidelberg, Germany}
\affiliation{GSI Helmholtzzentrum f{\"u}r Schwerionenforschung, 64291 Darmstadt, Germany}

\begin{abstract}
A fluid-dynamic approach to charm-quark diffusion in the quark-gluon plasma (QGP) is developed for the first time. Results for integrated yields and momentum distributions of charmed hadrons obtained with a fluid-dynamic description for the dynamics of the QGP coupled to an additional heavy-quark-antiquark current are shown. In addition to the thermodynamic Equation of State (EoS), this description uses a heavy-quark diffusion constant which we take from Lattice QCD calculations. The results describe the experimental data measured at the LHC at the center-of-mass energy of $\sqrt{s_{\rm NN}}$ = 5.02 TeV up to $p_{\rm T}\sim$ 4-5 GeV/$c$, showing that charm quarks undergo a very fast hydrodynamization in the medium created by ultrarelativistic heavy-ion collisions.

\end{abstract}
\maketitle

%%%%%%%%%%%%%%%%%%%%%%%%%%%%%%%%%%%%%%%%%%%%%%%%%%%%%%%%%%%%%%%%%%%%%%%%%%%%%%%%%%
%%%%%%%%%%%%%%%%%%%%%%%%%%%%%%%%%%%%%%%%%%%%%%%%%%%%%%%%%%%%%%%%%%%%%%%%%%%%%%%%%%
%%%%%%%%%%%%%%%%%%%%%%%%%%%%%%%%%%%%%%%%%%%%%%%%%%%%%%%%%%%%%%%%%%%%%%%%%%%%%%%%%%
\noindent \textit{Introduction.} Heavy quarks, i.e., charm and bottom, are produced in heavy-ion collisions (HICs) via hard partonic scattering processes. Due to their large mass and early production, they are suitable probes for studying the quark-gluon plasma (QGP). Their in-medium dynamics is tackled employing a Boltzmann/Langevin description in several transport models~\cite{Beraudo:2018tpr, Nahrgang:2014vza,Prado:2016szr, Song:2015ykw,Uphoff:2014hza, Cao:2016gvr,Ke:2018tsh, Scardina:2017ipo, He:2014cla,Rapp:2018qla}.
However, recent experimental measurements~\cite{ALICE:2020iug, ALICE:2020pvw} showed that J/$\psi$ and D mesons display a positive elliptic flow, suggesting an early local thermalization or ``hydrodynamization" of charm quarks within the QGP. The idea of charm thermalization was suggested by the Statistical Hadronization Model for charm~\cite{Andronic:2021erx} and supported by Lattice QCD (LQCD) calculations~\cite{Altenkort:2023oms}. In our recent work~\cite{Capellino:2022nvf}, the question of charm thermalization was addressed by studying the hydrodynamization time of charm quarks in the context of an expanding medium. It was shown that the time required for charm quark to hydrodynamize, and therefore to be included in the fluid-dynamic description of the QGP, is shorter than the typical expansion time scale of the medium. This result served as motivation to develop a fluid-dynamic description of charm quarks, which is the subject of the current work. We expect such a description to be relevant for the low transverse momentum ($p_{\rm T}$) region, as it is for the light-flavor particles. At high momentum, the path-length-dependent energy loss mechanisms, are more important in defining the shape of the $p_{\rm T}$ spectra. 

\noindent \textit{Fluid-dynamic equations.}
The fluid-dynamic equations to solve are mainly given by the system of equations
\begin{gather}
\label{eq:constmunu}
    \nabla_\mu T^{\mu\nu} = 0\,,\\
    \label{eq:conscurr}
    \nabla_\mu N^\mu = 0\,,
\end{gather}
which expresses the conservation of the energy-momentum tensor $T^{\mu\nu}$ and of an additional conserved current $N^\mu$. The latter is associated with conserving the number of charm-anticharm pairs~\cite{Capellino:2022nvf}.
The Landau frame is chosen such that $T^{\mu\nu}$ and $N^\mu$ can be decomposed as
\begin{gather}
    T^{\mu \nu} = (\epsilon+P)u^\mu u^\nu +\Delta^{\mu\nu}(P+\Pi)+\pi^{\mu\nu}\,,\\
    N^\mu = n u^\mu + \nu^\mu\,,
\end{gather}
where $\epsilon$, $P$, $u^\mu$, $\Pi$ and $\pi^{\mu\nu}$ are the energy density, thermodynamic pressure, fluid four-velocity, bulk viscous pressure, and shear-stress tensor of the fluid, respectively. The charm-quark fields are the heavy-quark density $n$ and the diffusion current $\nu^\mu$. The local temperature $T$ and the chemical potential to temperature ratio $\alpha$ are determined by the Landau matching conditions,
\begin{gather}
    \epsilon(T) \equiv \epsilon_{\rm equilibrium}(T)\,\\
    n(T,\alpha) \equiv n_{\rm equilibrium}(T,\alpha)\,.
\end{gather}
We assume that the energy density is approximately independent of the heavy-quark contribution, such that any energy density dependence on $\alpha$ is negligible. 
The thermal equilibrium heavy-quark density is taken to be one of the hadron-resonance gas, including all measured charm states (HRGc),
\begin{equation}
\label{eq:HRGc}
    n(T,\alpha) = \frac{T}{2\pi^2}\sum_{i\in \text{HRGc}} q_i M_i^2e^{q_i \alpha} K_2(M_i/T)\,,
\end{equation}
where $M_i$ is the mass of each charm hadron, and $q_i$ is its charm charge. 
The HRGc is expected to give the correct limit for the thermodynamics of the charm density at temperatures close to the phase transition. This relation is assumed to also hold at high temperatures. In the temperature regime reached by the fireball in most central collisions, the HRGc yields larger values (of about a factor 5) than the density of the free charm quarks. Nevertheless, due to the absence of first principle calculations for the Equation of State of charm quarks at physical QGP temperatures, we assume this relation to hold also at high temperatures. In the future, a more realistic Equation of State will be developed. 

The equations of motion for each of the dissipative currents in a second-order hydrodynamic formalism are solved,
\begin{align}
\label{eq:bulkeq}
    \tau_\Pi u^\mu \partial_\mu \Pi + \Pi +\zeta \nabla_\mu u^\mu &=0, 
    \\
    \nonumber
    P^{\mu\rho}_{\nu\sigma} \left[ \tau_\pi (u^\lambda \nabla_\lambda \pi^\sigma_\rho - 2 \pi^{\sigma \lambda} \omega_{\rho \lambda} 
    +\tfrac43 \nabla_\lambda u^\lambda \pi^{\sigma}_{\rho})+
    \right.&
    \\
    \left.
    +2\eta \nabla_\rho u^\sigma + \pi^\sigma_\rho \right]&=0, \label{eq:sheareq}\\
     \tau_n \Delta^\alpha_\beta u^\mu \nabla_\mu \nu^\beta + \nu^\alpha + \kappa_n \Delta^{\alpha\beta} \partial_\beta \alpha &=0\label{eq:nueq}
     \,,
\end{align}
where one defines the projector $P^{\mu\nu}_{\rho\sigma} = \frac12[\Delta^{\mu}_{\rho}\Delta^{\nu}_{\sigma}+\Delta^{\mu}_{\rho}\Delta^{\nu}_{\sigma}-\frac{2}{3}\Delta^{\mu}_{\rho}\Delta^{\nu}_{\sigma}]$ and the vorticity tensor $\omega^{\mu\nu}=(\nabla^\mu u^\nu-\nabla^\nu u^\mu)/2$. Here we introduced the transport coefficients for the bulk viscosity $\zeta$, shear viscosity $\eta$, and the heavy-quark diffusion coefficient $\kappa_n$, with the corresponding relaxation times $\tau_\Pi,\tau_\pi$ and $\tau_n$. The values of the viscosities are taken from Ref.~\cite{Devetak:2019lsk}, while the expression for the diffusion coefficient was derived in Ref.~\cite{Capellino:2022nvf}. We remark that $\kappa_n$ and $\tau_n$ are proportional to the heavy-quark spatial diffusion coefficient $D_s$.\\
The equations are solved in Bjorken coordinates assuming boost and azimuthal rotation invariance, restricting effectively to 1+1 dimensions.  
We organize the fluid fields for the QGP into a Nambu spinor $\Phi = (T,u^\mu,\pi^{\mu\nu},\Pi)$, which satisfies the hyperbolic equation of motion. We assume that none of these fields or transport coefficients depend on the heavy-quark variables. Eqs.~\eqref{eq:constmunu}, \eqref{eq:bulkeq} and~\eqref{eq:sheareq}, can be used to determine the time derivatives of the fluid fields explicitly.
Let us now consider another Nambu spinor including also the heavy-quark fields $\Xi = (T,u^\mu,\pi^{\mu\nu},\Pi,\alpha,\nu^\mu)$.
The new system of hyperbolic equations satisfied by $\Xi$ can be numerically solved by setting the fluid fields contained in $\Phi$ on shell. This is equivalent to neglecting the back reaction of the heavy-quark field on the fluid background evolution. More details regarding our approach to solve the fluid-dynamic equations, the numerical implementation and the validation of our fluid-dynamic framework can be found in the Appendix~\ref{sec:fequations},~\ref{sec:numerics} and~\ref{sec:gubser}.

\noindent \textit{Initial conditions for the fluid fields.}
The initial condition for the entropy density is computed with \trento~\cite{Moreland:2014oya} simulating Pb-Pb collisions at $\sqrt{s_{\rm NN}}=5.02$ TeV. 
The \trento parameters %\footnote{We set $w=0.5$~fm, $m=4$, $v=0.4$~fm, $p=0$ as in Ref.~\cite{Giacalone:2022hnz}. In addition, we set $k=1$ and $d=0.75$~fm, based on the outcome of the Bayesian analysis of Ref.~\cite{Nijs:2020ors}.} 
are set based in Refs.~\cite{Giacalone:2022hnz, Nijs:2020ors}; the \trento output is used as entropy density. The nucleon-nucleon cross section for Pb-Pb collisions at $\sqrt{s_{\rm NN}}=5.02$ TeV is taken from~\cite{ALICE:2022xir}, i.e.~$x=67.6$~mb. The nucleons in the Pb ion are sampled from a spherically symmetric Woods-Saxon distribution with radius $R=6.65$~fm and surface thickness $a=0.54$~fm. Using this set of parameters, the transverse density $T_{\rm R}(x,y)$ is generated for $1.5\cdot10^6$ minimum-bias collisions, among which the ones belonging to the 10\% most central are selected. The normalization of the \trento profile is computed by fixing the multiplicity of protons to the measured one~\cite{PhysRevC.101.044907}. 
The proton multiplicity in our calculation is obtained by employing a Cooper-Frye+FastReso approach at the end of the fluid-dynamic evolution as in Refs.~\cite{Devetak:2019lsk,Mazeliauskas:2018irt,Vermunt:2023fsr}. %However, the results do not significantly change if the normalization is fixed to reproduce the charged particles' multiplicity. 
In the future, when performing a Bayesian analysis to fit the experimental measurements, the normalization will be left as a free parameter of our model as in Ref.~\cite{Vermunt:2023fsr}.
The initial conditions for the temperature field are then obtained through the thermodynamics EoS described in Ref.~\cite{Floerchinger:2018pje}. Radial fluid velocity, shear-stress tensor components, and bulk viscous pressure are initialized at zero. 

\noindent \textit{Initial conditions for charm fields.}
The midrapidity density of charm quarks at the initialization time of the hydrodynamic evolution $\tau_0$ comes from the initial hard production, 
\begin{equation}
n^{Q\overline{Q}}_{\rm hard}(\tau_0,\vec x_\perp,y=0)=\frac{1}{\tau_0}\left.\frac{d^3N^{Q\overline{Q}}}{d\vec x_\perp dy}\right|_{y=0}
\,.\end{equation}
In the above expression, the $Q\overline{Q}$ rapidity distribution in nucleus-nucleus collisions is set by the pQCD $Q\overline{Q}$ cross-section
\begin{equation}
    \frac{dN^{Q\overline{Q}}}{dy}=\langle N_{\rm coll}\rangle \frac{1}{\sigma^{\rm in}}\frac{d\sigma^{Q\overline{Q}}}{dy}\,,
\end{equation}
where $\sigma^{\rm in}$ is the inelastic proton-proton cross-section and $\sigma^{Q\overline{Q}}$ is the hard production cross-section. 
The average number of collisions $N_{\rm coll}$ is computed with a Glauber model and depends on the impact parameter of the collision, providing:
\begin{equation}
    n^{Q\overline{Q}}_{\rm hard}(\tau_0,\vec x_\perp,y=0)=\frac{1}{\tau_0} n_{\rm coll}(\vec x_\perp) \frac{1}{\sigma^{\rm in}}\frac{d\sigma^{Q\bar Q}}{dy}\,,
\end{equation}
where $n_{\rm coll}$ is assumed to be distributed according to the fluid energy density $n_{\rm coll}\propto T^4$. As a future development, one could evaluate the radial distribution of binary collisions directly from \trento, not to neglect space-momentum correlations that are important for flow observables.
The integral of the density in the transverse plane provides the total number of heavy quarks to be conserved throughout the QGP evolution. As discussed in~\cite{Capellino:2022nvf}, we remark that the current associated with the number of heavy quark-antiquark pairs is \textit{accidentally} conserved. The heavy-quark mass is too large for them to be produced thermally throughout the QGP evolution; moreover, the annihilation rate of a $Q\overline{Q}$ pair is negligible within the lifetime of the plasma.

To fix at each point the initial value for $\alpha$ for the $Q\overline{Q}$ pair, 
\begin{equation}
n(T,\alpha)= n^{Q\overline{Q}}_{\rm hard}    
\end{equation}
Taking the central prediction by FONLL~\cite{Cacciari:2001td} for collisions at $\sqrt{s_{\rm NN}} =$ 5.02 TeV, one gets, at $y=0$, $d\sigma^{Q\overline{Q}}/dy=0.463$ mb, with $\sigma^{\rm in}=67.6$~mb~\cite{ALICE:2022xir}. 
At the beginning of the system evolution, the thermal distribution at zero chemical potential overshoots the density of charm quarks in the middle of the fireball. Therefore, $\alpha$ assumes negative values initially to match the hard production. This is not expected to happen at the fireball evolution's end, where the charm species' thermal abundance will be strongly suppressed.
The total multiplicity of $Q\overline{Q}$ pairs per unit of rapidity is given by the integrated density profile, e.g. at $\tau=\tau_0$,
\begin{equation}
    N^{Q\overline{Q}} = \tau_0 2\pi \int dr r n_{\rm hard}^{Q\overline{Q}}\,.
\end{equation}
In terms of fluid variables, due to the conservation of the charm current, the conserved charge is rewritten as,
\begin{equation}
    N^{Q\overline{Q}} = \int d^3x \sqrt{|g|} N^0(\vec{x}) = 2\pi \tau \int r ( n u^\tau + \nu^\tau) dr
\end{equation}
where $|g|$ is the determinant of the metric.
Besides the density, we can initialize the heavy-quark diffusion current. The assumed symmetries would allow a non-vanishing radial component, but we set it to zero in the absence of a more detailed initial state model. 

\begin{figure}
    \centering
    \includegraphics[width=0.45\textwidth]{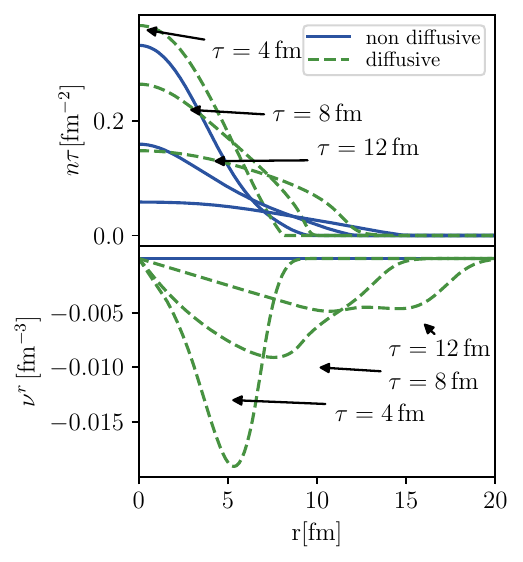}
    \caption{Charm density times the longitudinal proper time (upper panel) and diffusion current (lower panel) as a function of radius for different times. Solid lines correspond to an ideal hydrodynamic evolution, with $D_s=0$. Dashed lines correspond to a diffusive hydrodynamic evolution, with $2\pi D_s T$ taken from LQCD \cite{Altenkort:2023oms}.}
   \label{fig:density}
\end{figure}

\noindent \textit{Evolution of the fields.}
The initial conditions for the fields are set on a hypersurface at constant proper time $\tau_0 = 0.4$~fm. In Fig.~\ref{fig:density} (upper panel), the time evolution of the charm density times the longitudinal proper time as a function of the radial coordinate is reported for different values of $\tau$. This is shown for a non-diffusive ($D_s=0$) and temperature-dependent $D_s$ case obtained by linearly fitting results from LQCD calculations \cite{Altenkort:2023oms}. %It is assumed that $2\pi D_s T$ remains constant throughout the plasma evolution. 
As expected, the density becomes more diluted when the temperature decreases.
%\begin{figure}
%    \centering
%    \includegraphics{density.pdf}
%    \includegraphics{diffcurrent.pdf}
%    \caption{Charm density (upper panel) and diffusion current (lower panel) as a function of radius for different times. Solid lines correspond to an ideal hydrodynamic evolution, with $D_s=0$. Dashed lines correspond to a diffusive hydrodynamic evolution, with $2\pi D_s T = 1.5$ at all temperatures.}
%    \label{fig:density}
%\end{figure}
In the diffusive case, the density evolution is concurrent with developing the radial component of the diffusion current (Fig. \ref{fig:density}, lower panel). 
Its values are always negative, thus negatively contributing to the conserved current $N^\mu$. This results in a higher density $n$ in the diffusive case, as shown in Fig. \ref{fig:density}.
Comparing it to the equilibrium composition of the heavy-quark density $n$, one finds that the condition of $|\nu^r|\ll n$ is not satisfied in the entire radial region. This indicates that the out-of-equilibrium components of the heavy-quark distribution remain large throughout the evolution of the plasma. However, the magnitude of the diffusion current strongly depends on the spatial diffusion coefficient and its correspondent relaxation time. LQCD computations~\cite{Altenkort:2023oms} favor a fast hydrodynamization of charm quarks and, thus, a reduction of the out-of-equilibrium correction. 
Around freeze-out we decompose the single-particle distribution functions, $f_i = f_{i,{\rm eq}}+\delta f_i$, where the equilibrium part $f_{i,{\rm eq}}$ is given by the ideal gas distribution and $\delta f_i$ represents the out-of-equilibrium correction. In general, the $\delta f_i$ correction receives a contribution from all the dissipative stresses $\Pi$, $\pi^{\mu\nu}$ and $\nu^\mu$, such that
$
    \delta f_i = \delta f_{i,{\rm bulk}}+\delta f_{i,{\rm shear}}+\delta f_{i,{\rm diffusion}}\,.
$
In our case, the open-charm hadrons distribution function includes both light and heavy components. To properly describe it, one should derive its expression in a multi-species fluid setup. As for now, we neglect out-of-equilibrium corrections to the fluid variables at the freeze-out surface. In the future, we will address the inclusion of non-linear terms in the evolution equation for the dissipation current and the derivation of a more consistent expression of the total distribution function.

\begin{figure}
    \centering
    
\includegraphics[width=0.48\textwidth]{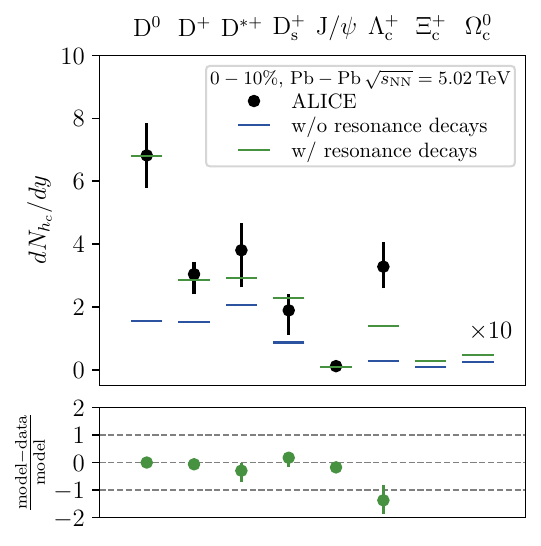}
    \caption{Charmed-hadron integrated yields with and without feed-down contributions from resonance decays and comparison with experimental data from the ALICE collaboration.}
    \label{fig:mult}
\end{figure}

\noindent \textit{Integrated yields.}
\label{sec:hadrons}
The charmed-hadron production is assumed to occur on a freeze-out hypersurface at a constant temperature. This chosen temperature is $T_{\rm fo}=156.5$ MeV~\cite{Andronic:2017pug,Andronic:2021erx}. The freeze-out hypersurface in the plane of Bjorken time $\tau$ and radius $r$ is parametrized by a parameter $\gamma \in (0,1)$. According to the Cooper-Frye prescription, a sudden decoupling is assumed at the freeze-out temperature, and the thermal momentum distribution of the particles is computed according to
\begin{align}
\label{eq:spectra}
\nonumber
    &\frac{dN_{h_c}}{p_T d\phi  dp_Tdy} = \frac{g_{h_c}}{(2\pi)^3}\int_{\Sigma_{\rm fo}} d\gamma d\phi dy \tau(\gamma)r(\gamma)
    \\
    \nonumber
    &e^{q\alpha}\left[\frac{\partial r}{\partial \gamma} m_T K_1\left(m_T \frac{u^r}{T}\right)I_0\left(p_T \frac{u^r}{T}\right)\right.\\
    &\left.-\frac{\partial \tau}{\partial \gamma} K_0\left(m_T \frac{u^r}{T}\right)I_1\left(p_T \frac{u^r}{T}\right)\right]\,,
\end{align}
where $g_{h_c}$ accounts for the degeneracy of the produced charmed hadron and $q$ accounts for the charm content of the hadron.
The total integrated yield $dN_{h_c}/dy$ per unit rapidity for charmed and anti-charmed hadrons is measured by integrating Eq.~\eqref{eq:spectra}.
The feed-down from resonance decays is calculated using the FastReso package~\cite{Mazeliauskas:2018irt}. The list of resonances is taken from the PDG~\cite{ParticleDataGroup:2022pth}. In Fig.~\ref{fig:mult}, the comparison between the obtained integrated yields and experimental measurements~\cite{ALICE:2021rxa, ALICE:2023gco,ALICE:2021bib, ALICE:2021kfc} is shown for the 0-10\% centrality interval. The yields and the $p_{\rm T}$ spectra correspond to the sum of particle and anti-particle divided by two, as reported by experiments. The $p_{\rm T}$ integration range is from 0 to 10~GeV/$c$.
These results are computed for $D_s=0$ since the integrated yield should not depend on the spatial diffusion coefficient. However, since out-of-equilibrium corrections to the single-particle distribution function at freeze-out are neglected, there can be a non-physical dependence of the yields on $D_s$.
While the relative abundance of each charmed-hadron species depends mainly on the mass of the hadron, the absolute value of the integrated yields strongly depends on the EoS for the charm density as a function of $T$ and $\alpha$. The HRGc as EoS is the most suitable choice to estimate the thermal production of the hadrons and resonances included in the HRGc. The role played by the resonance decays is then to reshuffle the relative abundance of the hadrons while keeping the total number of charm quarks fixed. 
The agreement between the model and the measurements is quantified in the lower panel of Fig.~\ref{fig:mult}. We observed that the mesons are compatible with the experimental uncertainties, computed as the sum in quadrature of the statistical and systematic uncertainties. A deviation of 2.4$\sigma$ is observed for the $\rm \Lambda_c^+$ baryons. This deviation might be caused by missing higher resonance states in the PDG~\cite{He:2019tik, He:2019vgs,Andronic:2021erx}. 
Due to the resonances decay, the yield of the $\rm D^0$ increases by a factor $4.3$, while the one of the $\rm \Lambda_c^+$ of a factor $5$. Another factor $2.3$ would be needed to reproduce the experimentally measured yield.
Estimates for the yields of the $\rm \Xi^+_c$ and $\rm \Omega^0_c$, whose experimental measurements are not yet available, are provided. Most likely, these values will underestimate the actual yields due to the lack of knowledge of higher resonance states. 
In other phenomenological models, the charmed-baryon enhancement is attributed to a recombination process between the heavy quark and light thermal partons~\cite{Minissale:2020bif, Plumari:2017ntm,Beraudo:2022dpz, Beraudo:2023nlq}.

\begin{figure}
    \centering
    \includegraphics[width=0.45\textwidth]{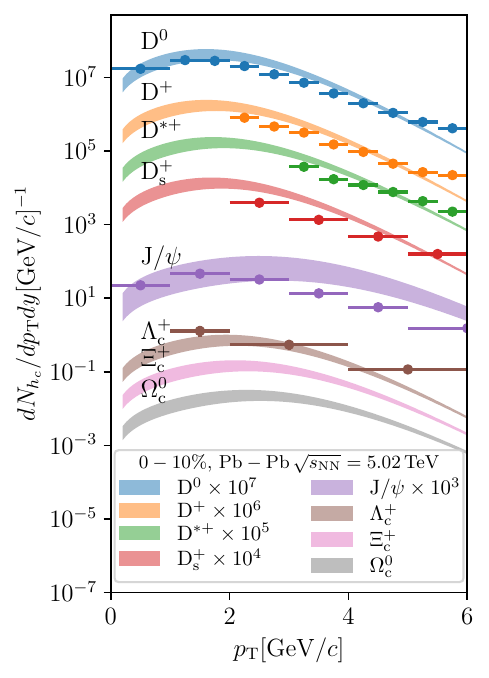}
    \caption{Results for the momentum distributions of $\rm D^0$, $\rm D^+$, $\rm D^{*+}$, $\rm D_s^+$, $\rm \Lambda_c^+$, and J/$\psi$ are shown in comparison with experimental measurements from the ALICE Collaboration~\cite{ALICE:2021rxa,ALICE:2023gco,ALICE:2021bib,ALICE:2021kfc}. Predictions for $\rm \Xi_c^0$ and $\rm \Omega_c^0$ baryon states, which have not been measured yet, are also shown.}
    \label{fig:spectra}
\end{figure}

\noindent \textit{Momentum distributions.}
In Fig.~\ref{fig:spectra}, the $p_{\rm T}$-differential spectra for the same hadron species are reported and compared with the experimental measurements~\cite{ALICE:2021rxa, ALICE:2023gco,ALICE:2021bib, ALICE:2021kfc}. A ratio plot with the data to model comparison can be found in Appendix~\ref{sec:mom_distr}.
The bands correspond to a spread of the input value of the spatial diffusion coefficient $D_s$ going from a non-diffusive case ($D_s = 0$) to a temperature-dependent $2\pi D_s T$~\cite{Altenkort:2023oms}. The fluid-dynamic description seems to capture the physics of D mesons up to $p_T\sim$~4--5~GeV/$c$. This implies that, even if the charm does not move collectively with the rest of the fluid in the early stage of the evolution, it relaxes to the same radial flow of the QGP before the freeze-out occurs. As observed for the integrated yield, the $\Lambda_c^+$ calculation underestimates the experimental measurement. The J/$\psi$ $p_{\rm T}$ distribution describes the experimental measurements for $p_{\rm T} < 3$ GeV/$c$, while it overpredicts the yield for higher $p_{\rm T}$. This discrepancy for $p_{\rm T} >$ 3 GeV/$c$ might be attributed to the dominant contribution from primordial J/$\psi$, which is not accounted for in our model since it is not expected to reach thermal equilibrium~\cite{Zhao:2007hh,He:2021zej,Song:2017phm}, but is mainly sensitive to path-length-dependent effects, like survival probability and energy loss.
It is also important to note that the experimental measurements consist of J/$\psi$ directly produced in the collisions plus the contribution from beauty hadron decays. Including the out-of-equilibrium corrections in the model at the freeze-out surface will influence the shape of the momentum distributions. They would modify the spectra at intermediate/high $p_{\rm T}$. When adequately included, we do not expect such a strong dependence on $D_s$ in the spectra but rather only a tilt in the momentum distribution. 
A further remark regards the dependence of the final momentum distribution on the initial conditions for the charm fields. In particular, a broader initial distribution for the charm density results in a larger average $p_T$ at freeze-out. A more thorough study of the charm initial conditions will improve the description of the transverse momentum distribution of the charm hadrons, without of course impacting the results for the integrated yields.

\noindent \textit{Conclusions.}
A fluid-dynamic description of the charm quark is developed for the first time, unveiling that low-$p_{\rm T}$ charm quarks undergo a very fast hydrodynamization in the QGP created during ultrarelativistic heavy-ion collisions. The developed model describes the charmed-hadron yield and the $p_{\rm T}$-differential distribution up to $p_{\rm T}\sim$ 4--5 GeV/$c$. 
The calculations are carried out for a non-diffusive case and a 
temperature-dependent $D_s$. The derivation of out-of-equilibrium corrections in a multi-species setup will be addressed. Additional constraints on the spatial diffusion coefficient will be set in future works via a Bayesian analysis using both $p_{\rm T}$-differential spectra and anisotropic flow coefficients. In addition, this study paves the way for a fluid-dynamic description of the beauty quark, which, despite its larger mass, might still reach a partial local equilibrium allowing further constraining QGP parameters using heavy quarks as probes.

\textit{Acknowledgment}
The authors wish to thank A. Andronic and A. Mazeliauskas for the useful discussions about the resonance decay list and Andrea Beraudo for a careful review of the manuscript.
This work is part of and supported by the DFG Collaborative Research Centre "SFB 1225 (ISOQUANT)".
A.D. is partially supported by the Netherlands Organisation for Scientific Research (NWO) under the grant 19DRDN011, VI.Veni.192.039.

\bibliography{bibliography}

\appendix

\section{Fluid-dynamic equations}
\label{sec:fequations}

The equations are solved effectively in 1+1 dimensions with Bjorken coordinates $(r,\tau)$ supplemented by azimuthal angle $\phi$ and rapidity $\eta$. The metric tensor is defined as $g_{\mu \nu} = \mathrm{diag}(-1,1,r^2,\tau^2)$. 
%%%%%%%%%%%%%%%%%%%%%%%%%%%%%%%%%
The independent components of the fluid fields in the azimuthally symmetric and boost-invariant case are, for the energy-momentum tensor $T^{\mu\nu}$ 
\begin{equation}
T,u^r,\pi^{\eta}_{\eta},\pi^{\phi}_{\phi},\Pi,
\end{equation}
where $u^2=-1$ and $\pi^{\mu\nu}$ is a symmetric, traceless tensor transverse to the four-velocity.
The independent charm fields are
\begin{equation}
    \alpha, \nu^r,
\end{equation}
where the diffusion current is orthogonal to the fluid velocity $u\cdot \nu=0$, and $\alpha$ is the conjugate variable of the density $n$, i.e., $\alpha =\mu/T$.
%%%%%%%%%%%%%%%%%%%%%%%%%%%%%%%%%
A generic Nambu spinor, whose components are the background fluid fields $\Phi_{\rm bg} = (T,u^\mu,\pi^{\eta}_{\eta},\pi^{\phi}_{\phi},\Pi)$, is considered. 
The general hyperbolic equations for the background fluid field can be written as
\begin{align}
\label{eq:hyp_bg}
&A_{\rm bg}(r,\tau,\Phi_{\rm bg}) \partial_\tau  \Phi_{\rm bg} +\\\nonumber
&B_{\rm bg}(r,\tau,\Phi_{\rm bg})\partial_r \Phi_{\rm bg}
=
S_{\rm bg}(r,\tau,\Phi_{\rm bg})\,,
\end{align}
where $A_{\rm bg}$ and $B_{\rm bg}$ are $5\times 5$ matrices and $S_{\rm bg}$ is a source term vector depending non-linearly on the fluid fields $\Phi_{\rm bg}$. This equation is used to derive the expressions for the time derivatives of the fields in $\Phi_{\rm bg}$,
\begin{equation}
\label{eq:onshellbg}
    \partial_\tau \Phi_{\rm bg} =-A_{\rm bg}^{-1}B_{\rm bg}\partial_r \Phi_{\rm bg}+
    A_{\rm bg}^{-1}S _{\rm bg}\,.
\end{equation}
We can now define another Nambu spinor $\Phi_{\rm HQ}=(\alpha,\nu^r)$. The equation of motion for $\Phi_{\rm HQ}$ is given by
\begin{align}
\label{eq:hyp_charm}
    &A_{\rm HQ} \partial_{\tau}\Phi_{\rm HQ} + \\ 
    \nonumber&B_{\rm HQ}  \partial_{r}\Phi_{\rm HQ} + C_{\rm HQ} \partial_{\tau}\Phi_{\rm bg} + D_{\rm HQ}  \partial_{r}\Phi_{\rm bg}  =S_{\rm HQ} \,, 
\end{align}
where the $2\times2$ matrices $A_{\rm HQ}$, $B_{\rm HQ}$, $C_{\rm HQ}$, $D_{\rm HQ}$ and the two-component vector $S_{\rm HQ}$ are generic non-linear functions of the fluid fields $\Phi_{\rm bg}$ and the heavy quarks variables  $\Phi_{\rm HQ}$. 
We substitute the equation of motion Eq.~\eqref{eq:onshellbg} in Eq.~\eqref{eq:hyp_charm}, leading to 
\begin{align}
\label{eq:hyp_charm_onshell}
    &A_{\rm HQ} \partial_{\tau}\Phi_{\rm HQ} + B_{\rm HQ}  \partial_{r}\Phi_{\rm HQ}+\\
    &\nonumber(D_{\rm HQ} - C_{\rm HQ} A_{\rm bg}^{-1}B_{\rm bg}  )  \partial_{r}\Phi_{\rm bg}  =S_{\rm HQ} -C_{\rm HQ}A_{\rm bg}^{-1}S _{\rm bg} .
\end{align}
In this formulation, the evolution of the background $\Phi_{\rm bg}$ is not influenced by the dynamics of the diffusion current and density in $\Phi_{\rm HQ}$. The equations of motion read
\begin{align}
\label{eq:allfieldsbg}
&\begin{pmatrix}
     & A_{\rm bg} & 0\\
     & 0          &  A_{\rm HQ}
    \end{pmatrix}
   \partial_\tau 
    \begin{pmatrix}
    \Phi_{\rm bg}\\
    \Phi_{\rm HQ}
\end{pmatrix} 
+\\\nonumber
&\begin{pmatrix}
     & B_{\rm bg} & 0\\
     & B_{\rm mix} & B_{\rm HQ}
\end{pmatrix}
\partial_r 
    \begin{pmatrix}
    \Phi_{\rm bg}\\
    \Phi_{\rm HQ}
\end{pmatrix} 
=
 \begin{pmatrix}
    S_{\rm bg}  \\
    \widetilde S_{\rm HQ}
\end{pmatrix}
\end{align}
where the mixing matrix is given by $B_{\rm mix}= D_{\rm HQ} - C_{\rm HQ} A_{\rm bg}^{-1}B_{\rm bg}$ and $\widetilde S_{\rm HQ}=S_{\rm HQ} -C_{\rm HQ}A_{\rm bg}^{-1}S _{\rm bg} $.
Using that the equations are hyperbolic, and therefore the matrix of the time derivative $\begin{pmatrix}
     & A_{\rm bg} & 0\\
     & 0          &  A_{\rm HQ}
    \end{pmatrix}$ is invertible, the equations of motion can be written explicitly as
\begin{align}
   &\partial_\tau 
    \begin{pmatrix}
    \Phi_{\rm bg}\\
    \Phi_{\rm HQ}
\end{pmatrix} 
+\\ &\nonumber
\begin{pmatrix}
     & A^{-1}_{\rm bg}B_{\rm bg} & 0\\
     & A^{-1}_{\rm HQ}B_{\rm mix} & A^{-1}_{\rm HQ}B_{\rm HQ}
\end{pmatrix}
\partial_r 
    \begin{pmatrix}
    \Phi_{\rm bg}\\
    \Phi_{\rm HQ}
\end{pmatrix} 
=
\begin{pmatrix}
    A^{-1}_{\rm bg}S_{\rm bg}  \\
    A^{-1}_{\rm HQ}\widetilde S_{\rm HQ}
\end{pmatrix}\,.
\end{align}

\section{Numerical scheme}
\label{sec:numerics}

The equations of motion for relativistic fluid dynamics with the conservation of a heavy-charm pair current are hyperbolic equations of motion due to the inclusion of the evolution equation of the dissipative currents $\pi^{\mu\nu},\Pi$ and $\nu^\mu$. Schematically, the equations can be written as quasi-linear partial differential equations (PDEs).
We will restrict ourselves to discussing the equations in one spatial dimension for simplicity. However, the extension to a higher number of dimensions is trivial. 
We consider a collection of independent variables called $\phi$, whose equations of motion are  
\begin{equation}
\label{eq:non-conservative}
\partial_t \phi + A(\phi)\partial_x\phi+ S(\phi)=0\,,
\end{equation}
where $A(\phi)$ is a matrix in field space that depends non-linearly on the fields themselves, and $S(\phi)$ is a vector containing the source term in the equation. 
Usually, the numerical solutions of the fluid dynamic equations are discussed in a conservative form since the ideal limit of the equations is the divergence of a current -- typically the energy-momentum and particle density current. Let
\begin{equation*}
    \nabla_{\mu} \mathcal{J}^{\mu}=0
\end{equation*}
be the conservation equation, where $ \mathcal{J}^{\mu}$ represents generically the conserved current. 
However, including the dynamics of the dissipative current like diffusion and shear/bulk viscous stress spoils this property for Israel-Stewart-M\"uller theory~\cite{14-moment}. For this type of theory, the equations are non-conservative by construction, and it is impossible to cast them in a conservative form. 
In the relativistic viscous fluid dynamic literature, 
the equations are solved with a splitting algorithm:
First, solve using a finite volume conservative scheme, then correct the intermediate solution using a central approximation of the dissipative equations, as in the so-called SHASTA algorithm~\cite{Boris:1973tjt}, or some variations of it like KT~\cite{kurganov2000new}. 
This type of algorithm performs well if the dissipative currents are minor corrections to the ideal step and do not modify the ideal evolution substantially.  
However, this is not always the case, especially when the system is far from the ideal approximation, meaning the non-equilibrium effects are important. 
In this work, we implement a different strategy. Instead of using the ideal-viscous splitting, we solve the equations together as a quasi-linear system of PDEs. 
The naive discretization of equations like Eq.~\eqref{eq:non-conservative} can be obtained by replacing the first derivative with its central approximation. 
Denoting $x_i$ the central position of a cell of size $\Delta x$, the 
the central derivative approximation is 
\begin{equation}
    \partial_x \phi|_{x_i} \simeq 
    \frac{1}{2 \Delta x} (\phi_{i+1}-\phi_{i-1} )\,,
\end{equation}
where $\phi_i= \phi(x_i)$.
The semi-discretized version of the equations is 
\begin{equation}
  \partial_t \phi_i + A(\phi_{i})\partial_x\phi|_{x_i}+ S(\phi_i)=0.   
\end{equation}
This naive discretization, however, is unstable since there is no dissipation mechanism in the discretization to reduce the high-frequency mode of the discretized solution.
The physical motivation for this instability can be understood considering the nature of the PDE. 
The system of hyperbolic equations is a collection of propagating waves that interact non-linearly and with a non-constant velocity. 
The waves are usually (except in simple cases) a complicated combination of the primary variables $\phi$, defined as the left eigenvector of the matrix $A(\phi)$. 
The eigenvalue is characteristic of the hyperbolic PDE and represents how fast the wave propagates. 
Each of the waves propagates at a different speed and direction. In a one-dimensional case, there will be right- and left-moving waves. To have a stable discretization, the numerical derivative should respect -- up to some degree of accuracy -- the direction of propagation of the different waves. 
If a wave is right-moving, the correct derivative discretization should involve only points in the past of the wave -- i.e. on its left -- and vice versa. This mechanism is called \textit{upwinding}~\cite{osher1982upwind}. Therefore, the central approximation of the first derivative goes against this principle since it does not distinguish the direction of propagation of the waves.  

A natural solution is to separate right-moving and left-moving waves and discretize them accordingly. By calling $\lambda_i$ the eigenvalues, one can separate them into positive and negative ones ($\lambda_i^{+}$ and $\lambda_i^{-}$, respectively)\footnote{For hyperbolic systems of partial differential equations, it is always possible to left-diagonalize the characteristic matrix and the corresponding eigenvalues are real.}, 
\begin{equation}
A^{+} = U \begin{bmatrix}
    \lambda^{+}_{1} & & \\
    & \ddots & \\
    & & 0
  \end{bmatrix} U^{-1 },\quad A^{-} = U \begin{bmatrix}
    0 & & \\
    & \ddots & \\
    & & \lambda^{-}_{1}
  \end{bmatrix} U^{-1 }\,.
\end{equation}
Each matrix has only information about the left and right propagating waves, respectively. 
With this construction, it is then easy (in principle) to construct an upwinding discretization as, 
\begin{equation}
\label{eq:discretization}
    \partial_t \phi_i + A^{+}(\phi_{i})\partial_x\phi|^{-}_{x_i}
    +A^{-}(\phi_{i})\partial_x\phi|^{+}_{x_i}
    + S(\phi_i)=0\,,
\end{equation}
where the derivatives are taken from the left or the right, respectively, 
\begin{equation}
    \phi|^{-}_{x_i}= \frac{1}{\Delta x }(\phi_i - \phi_{i-1}),\quad \phi|^{+}_{x_i}= \frac{1}{\Delta x }(\phi_{i+1} - \phi_{i})\,.
\end{equation}
The proposed discretization is sometimes called the flux-splitting technique and was already introduced in~\cite{osher1982upwind,YAN2011232,inbook}.
The drawback of this scheme is that it relies on the complete knowledge of the spectrum of the characteristic matrix. 
Only in a few cases is this achievable due to the complexity of the non-linearities of the characteristic matrix $A$. 

The discretization reported in Eq.~\eqref{eq:discretization} can be expressed in terms of the absolute value of the matrix $A$, 
\begin{align}
|A|= A^{+}-A^{-}\,,
\end{align}
such that 
\begin{align}
A^{+}=\frac{1}{2}(A + |A| )\,,\quad
A^{-}=\frac{1}{2}(A - |A| )\,.
\end{align}
Therefore, Eq. \eqref{eq:discretization} becomes
\begin{align}
   & \partial_t \phi_i +
    \frac12 A(\partial_x\phi|^{-}_{x_i}+\partial_x\phi|^{+}_{x_i})
    + \\
   & \nonumber \frac12|A| (\partial_x\phi|^{-}_{x_i}-\partial_x\phi|^{+}_{x_i})
    + S(\phi_i)=0.  
\end{align}
The derivative operators now become 
\begin{align}
  \frac12  (\partial_x\phi|^{-}_{x_i}+\partial_x\phi|^{+}_{x_i})=
  \frac{1}{2\Delta x }(\phi_{i+1} - \phi_{i-1})= \partial_x\phi|_{x_i}\,,\\
  \partial_x\phi|^{-}_{x_i}-\partial_x\phi|^{+}_{x_i}=
  \frac{1}{\Delta x }(\phi_{i+1} + \phi_{i-1}-2\phi_{i} )=\Delta x \partial^2_x\phi|_{x_i}\,,
\end{align}
leading to a discretized equation of the form
\begin{align}
    \partial_t \phi_i &+
    A\partial_x\phi|_{x_i}
    - \frac12|A|\Delta x \partial^2_x\phi|_{x_i}
    + S(\phi_i)=0\,.  
\end{align}
The extra contribution introduced to upwind the derivative acts like a 
viscous terms into the equation, with an amplitude proportional to the lattice spacing $\Delta x$. 

A standard approximation for the absolute value of the matrix is $|A|=\lambda I$ where $\lambda=\max(|\lambda_i|)$, which is the fastest characteristic speed in the system. Under this assumption, the scheme can be considered a non-conservative version of the Lax-Friedrichs scheme. 
However, this requires knowledge of the characteristic structure, which is possible only for exceptional cases. 

An appealing alternative is to approximate $|A|$ with a suitable expansion, as discussed in~\cite{absolute,inbook}. 
Among the possible expansion choices that one can make, the simplest is a polynomial approximation around $\max(|\lambda_i|)=1$, 
\begin{equation}
    |A|\simeq \frac12(I+A^2 ) + \mathcal{O}(A^4)\,,
\end{equation}
assuming that all the $|\lambda_i|<1$, such that the fastest wave speeds are modified correctly. 
Different and more performing possibilities are Chebyshev polynomials and rational functions. However, in this work, we restrict ourselves to the simplest choice and will study these possibilities in the future.

For the evolution, we use the explicit Runge-Kutta with adaptive time-step as described in~\cite{Julia-2017,rackauckas2017differentialequations} and with the Proportional-Integral-Derivative (PID) controller as described in~\cite{10.1145/641876.641877,SODERLIND2006225,ranocha2022optimized}. For the integration on the freeze-out surface we used \cite{HCubature,Genz1980,quadgk,ApproxFun.jl-2014}.

\section{Validation against Gubser flow}
\label{sec:gubser}
\begin{figure}

    \centering
    \includegraphics[width = 0.48\textwidth]{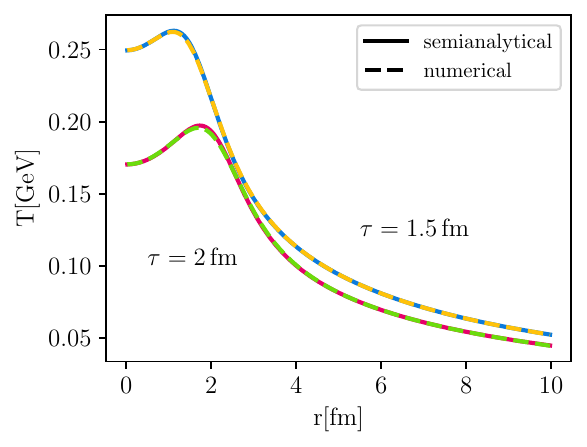}
    \includegraphics[width = 0.48\textwidth]{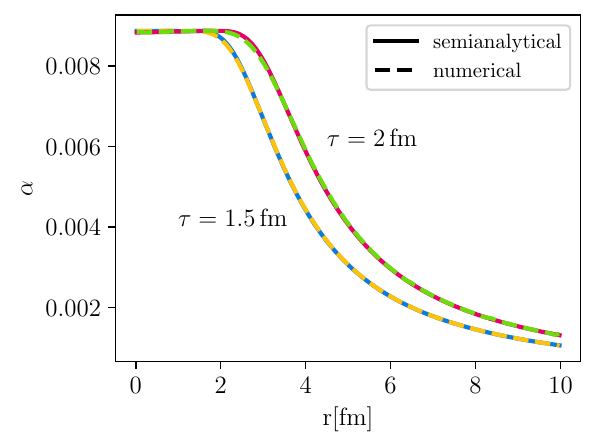}

    \caption{Temperature (upper panel) and chemical potential to temperature ratio $\alpha$ (lower panel) as a function of radius $r$ at Bjorken times $\tau$ = 1.5 fm/c and $\tau$ = 2 fm/c. The solid lines correspond to the semianalytic Gubser solution, while the dashed lines are the numerical result with N = 200 discretization points. We have here chosen the maximal radius to be 10 fm.}
\label{fig:gubser_bulk}   
\end{figure}

Comparing it against a known analytic (or semi-analytic) solution is useful to verify and validate the numerical scheme. For Israel-Stewart-type theories, such a solution with azimuthal rotation symmetry, longitudinal boost invariance, and an additional conformal symmetry has been found by Gubser~\cite{Gubser:2010ui}. For symmetry reasons, the evolution of the diffusion current in this setup is trivial. So we will leave it out of the discussion in the rest of this section. The set of equations for the evolution of temperature, fluid velocity, shear stress, number density in the presence of a conformal symmetry reads,
\begin{gather}
    \frac{u^\lambda\nabla_\lambda T}{T}+\frac{\nabla_\mu u^\mu}{3}+\frac{\pi^{\mu\nu}\sigma_{\mu \nu}}{3sT}=0\,,\\
    u^\lambda\nabla_\lambda u^\mu +\frac{\Delta^\mu_\lambda \nabla^\lambda T}{T} + \frac{\Delta^\mu_\lambda \nabla_\alpha \pi^{\alpha \lambda}}{sT}=0\,,\\
    \frac{\tau_\pi}{sT}(\Delta^\mu_\alpha \Delta^\nu_\beta u^\lambda \nabla_\lambda \pi^{\alpha \beta}+\frac{4}{3}\nabla_\lambda u^\lambda \pi^{\mu\nu})+\frac{\pi^{\mu\nu}}{sT}=-\frac{2\eta}{sT}\sigma^{\mu\nu}\,,\\
    u^\lambda\nabla_\lambda n = -n \theta - \nabla_\mu \nu^\mu\,,
\end{gather}
where $\theta=2\tanh\rho$ is the scalar expansion rate for Gubser flow. 
In de Sitter space, by applying the Gubser flow profile $\hat u^\mu=(1,0,0,0)$, the equations read,
\begin{gather}
    \frac{1}{\hat T}\partial_\rho \hat T + \frac{2}{3}\tanh \rho = \frac{1}{3} \bar \pi^\eta_\eta \tanh \rho\,,\\
    \frac{c}{\hat T} \frac{\eta}{\hat s} \left[\partial_\rho \bar \pi^\eta_\eta+\frac{4}{3}(\bar \pi^\eta_\eta)^2\tanh \rho\right] + \bar \pi^\eta_\eta = \frac{4}{3}\frac{\eta}{\hat s \hat T} \tanh \rho\,,\\
    \partial_\rho \hat{n} + 2\tanh \rho \hat{n} = 0\,,
\end{gather}
where $\rho$ is the Gubser conformal time variable and $\bar \pi^{\mu\nu}=\pi^{\mu\nu}/(\hat s \hat T)$. 
The transformation rules to obtain the fluid variables in Milne coordinates are given by
\begin{gather}
    T(\tau,r)=\hat T(\rho(\tau,r))/\tau\,,\\
    u_\mu (\tau,r)=\tau \frac{\partial \hat x^\nu}{\partial \hat x^\mu}\hat{u}_\nu(\rho(\tau,r))\,,\\
    \pi_{\mu\nu} (\tau,r)=\frac{1}{\tau^2} \frac{\partial \hat x^\alpha}{\partial \hat x^\mu}\frac{\partial \hat x^\beta}{\partial \hat x^\mu}\hat\pi_{\alpha \beta}(\rho(\tau,r))\,,\\
    n(\tau,r) = \frac{1}{\tau^3}\hat n(\rho(\tau,r))\,.
\end{gather}
The conformal Equation of State at finite $\alpha$ can be written as
\begin{gather}
    e = 3p\,,\quad
    s = h(\alpha) T^3\,,\quad
    n \equiv g(\alpha)\alpha T^3\,,
\end{gather}
where one defines the dimensionless coefficients
\begin{gather}
    f = 3 p_0+\frac{N_f}{6}\alpha^2+\frac{Nf}{108\pi^2}\alpha^4\,,\\
    h = 4p_0 +\frac{N_f}{9}\alpha^2\,,\\
    g = \frac{N_f}{9}\alpha+\frac{N_f}{81 \pi^2}\alpha^2\,.
\end{gather}
Here we use $p_0 = (16+10.5 N_f)\pi^2/90$ and the number of flavors $N_f = 2.5$. In this setup, the equations for the charge current are decoupled from the rest of the system.
In Fig.~\ref{fig:gubser_bulk} the comparison between the semi-analytical solution by Gubser and the one obtained numerically is presented. The initialization time is $\tau_0 = 1$ fm, the shear viscosity to entropy ratio is 0.2; the shear relaxation time is $\tau_S = 5\eta/(sT)$. 
The overall agreement is good for all fields in the full radial range. 

\begin{figure}
    \centering
    \includegraphics[width = 0.48\textwidth]{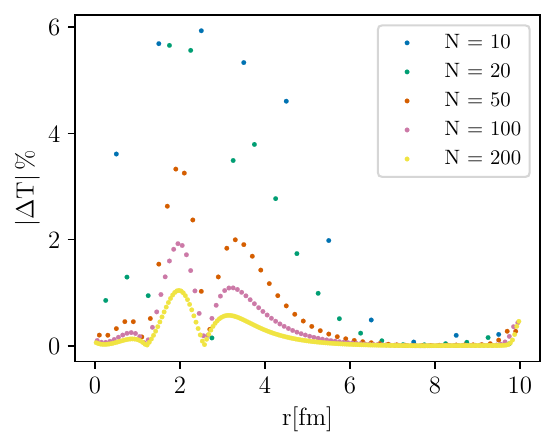}
    \caption{Percent deviation of the absolute value of $\Delta T = T_{\rm numerical}/T_{\rm semianalytical}-1$ of the temperature at $\tau = 2$ fm as a function of radius and the number of discretization point $N$. The numerical solution converges to the semi-analytical one as the number of points increases.}
    \label{fig:deltaT}
\end{figure}

In Fig.~\ref{fig:deltaT}, the percent deviation of the numerical solution for the temperature field with respect to Gubser's solution is shown for different numbers of discretization points at $\tau = 2$ fm. As one can see, the finer the spatial grid is, the smaller the deviation. In particular, the deviation around 2 fm, corresponding to the maximum of the temperature profile, is progressively suppressed. 

\section{Details on the results of charm-hadron momentum distributions}
\label{sec:mom_distr}
\begin{figure}
    \centering
    \includegraphics[scale = 1]{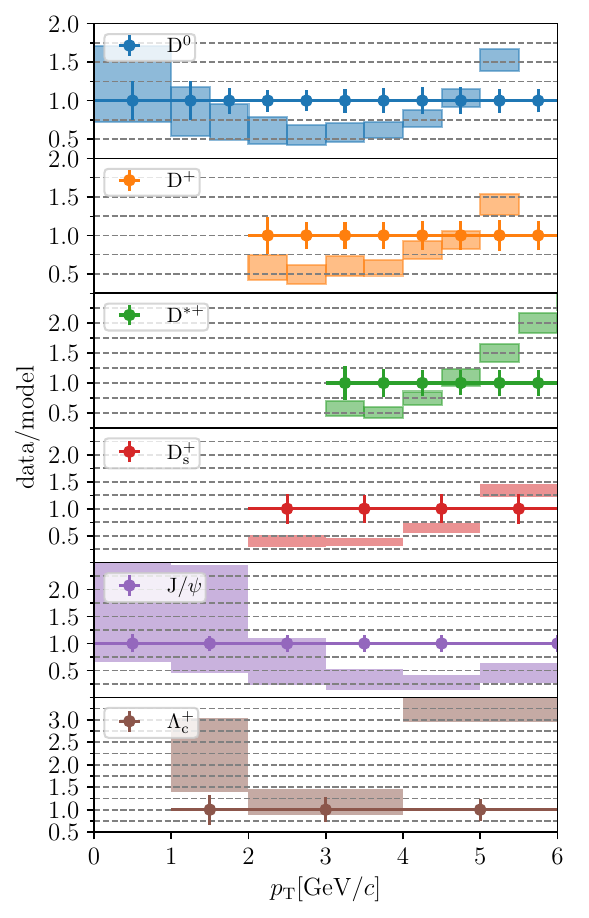}
    \caption{Data-to-model ratios for $\rm D^0$, $\rm D^+$, $\rm D^{*+}$, $\rm D_s^+$, $\rm \Lambda_c^+$, and J/$\psi$ momentum distributions. Experimental measurements are taken from~\cite{ALICE:2021rxa,ALICE:2023gco,ALICE:2021bib,ALICE:2021kfc}.}
    \label{fig:mom_distr}
\end{figure}

In Fig. \ref{fig:mom_distr} the results for the ratio between the experimental measurements of charm-hadron momentum distributions and the results from our fluid-dynamic model are shown. The bands correspond to a spread of the input value of the spatial diffusion coefficient $D_s$ going from a non-diffusive case ($D_s = 0$) to a temperature-dependent $2\pi D_s T$ obtained by linearly fitting results from LQCD calculations~\cite{Altenkort:2023oms}. The fluid-dynamic descriptions captures the behavior to ${\rm D^0}$ and ${\rm J/\Psi}$ up to $p_{\rm T}\sim 2$ GeV. At intermediate transverse momentum, our calculation for the ${\rm D}$ mesons deviates of 25$\%$ from the experimental measurements for the $D_s$ = 0 case. A larger deviation is hereby observed for ${\rm J/\Psi}$ attributed to the dominant contribution from primordial J/$\psi$, which is not accounted for in our model since it is not expected to reach thermal equilibrium~\cite{Zhao:2007hh,He:2021zej,Song:2017phm}, but is mainly sensitive to path-length-dependent effects, like survival probability and energy loss.
As observed for the integrated yield, the $\Lambda_c^+$ calculation underestimates the experimental measurement.  This deviation might be caused by missing higher resonance states in the PDG~\cite{He:2019tik, He:2019vgs,Andronic:2021erx}. 
At $p_{\rm T}$ larger than 5 GeV, the fluid-dynamic model seems no longer applicable since it's not able to capture the behavior of the particle spectra.

A further remark regards the dependence of the final momentum distribution on the initial conditions for the charm fields. In particular, a broader initial distribution for the charm density results in a larger average $p_T$ at freeze-out. A more thorough study of the charm initial conditions will improve the description of the transverse momentum distribution of the charm hadrons, without of course impacting the results for the integrated yields.

\end{document}